\documentclass{jfm}

\usepackage{graphicx}
\usepackage{amssymb}
\usepackage{amsfonts}
\usepackage{natbib}
\usepackage{pstool}
\usepackage{xcolor}
\usepackage{verbatim}
\usepackage{pifont}
\usepackage{psfrag}
\usepackage{amsmath}
\usepackage{appendix}
\usepackage{mathtools}

\newcommand{\ajs}[1]{{{\color{red} #1}}}

\title[Reynolds stress scaling in the near-wall region of wall-bounded flows]{Reynolds stress scaling in the  near-wall region of wall-bounded flows}

\author[Smits, Hultmark, Lee, Pirozzoli, Wu]%
{Alexander J. Smits$^1$, Marcus Hultmark$^1$, Myoungkyu Lee$^2$,\\ Sergio Pirozzoli$^3$ and Xiaohua Wu$^4$}

\affiliation{$^1$Department of Mechanical and Aerospace Engineering, 
Princeton University, \break
Princeton, NJ 08544, USA\\[\affilskip]
$^2$Sandia National Laboratories, 
Livermore, CA 94551, USA\\[\affilskip]
$^3$Dipartimento di Ingegneria Meccanica e Aerospaziale, Universit\`{a} di Roma ``La Sapienza,'' \break
00184 Roma, Italy\\[\affilskip]
$^4$Department of Mechanical and Aerospace Engineering, 
Royal Military College of Canada, \break
Kingston, Ontario, Canada
K7K 7B4\\[\affilskip]
}

\pubyear{2021}
\volume{}
\pagerange{}

\begin{document}

\maketitle

\begin{abstract}

A new scaling is derived that yields a Reynolds number independent profile for all components of the Reynolds stress in the near-wall region of wall bounded flows, including channel, pipe and boundary layer flows.  The scaling demonstrates the important role played by the wall shear stress fluctuations and how the large eddies determine the Reynolds number dependence of the near-wall turbulence behavior. 
\end{abstract}

\section{Introduction}

Here, we examine the near-wall scaling behavior of canonical turbulent flows on smooth surfaces.   These flows include two-dimensional zero-pressure gradient boundary layers, and fully developed pipe and channel flows.  The focus is on the region $y^+<100$, which includes the peaks in the streamwise and spanwise turbulent stresses.  Here, $y$ is the distance from the wall, and the superscript $^+$ denotes non-dimensionalization using the fluid kinematic viscosity  $\nu$ and the friction velocity $u_\tau=\sqrt{\tau_w/\rho}$, where $\tau_w$ is the mean wall shear stress and $\rho$ is the fluid density.   

For isothermal, incompressible flow, it is commonly assumed that for the region close to the wall
$$[U_i,\overline{u_i u_j}] = f(y, u_\tau, \nu, \delta),$$
where $U_i$ and $u_i$ are the mean and fluctuating velocities in the $i$th direction.  The overbar denotes time averaging, and the outer length scale $\delta$ is, as appropriate, the boundary layer thickness, the pipe radius, or the channel half-height.  That is,
\begin{equation}
[U_i^+,\overline{({u_i u_j})^+}] = f(y^+, Re_\tau).
\label{dim_analysis}
\end{equation}
where the friction Reynolds number $Re_\tau=\delta u_\tau/\nu$. 
\begin{figure}
\centering
\includegraphics[width=\textwidth]{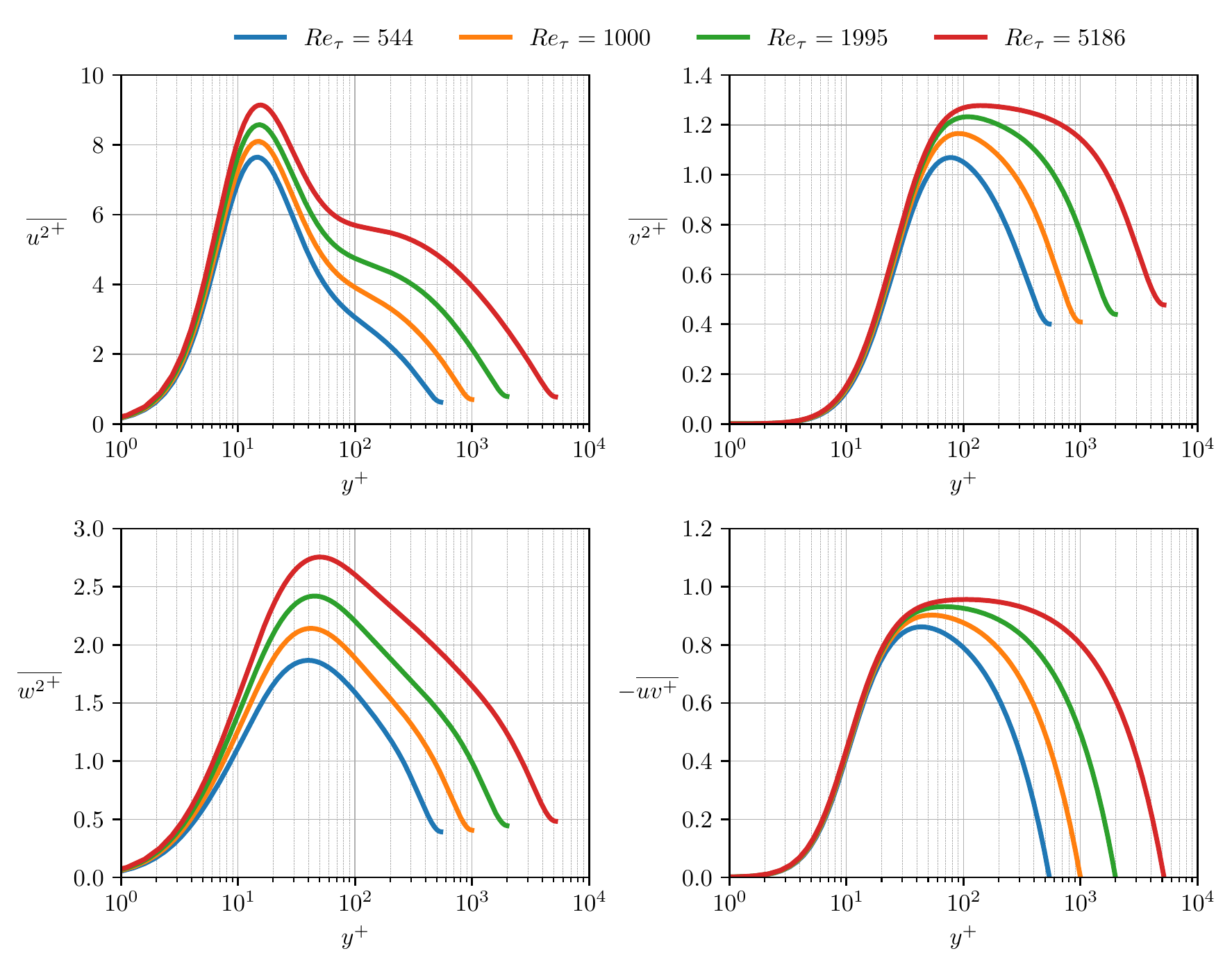} 
\caption{ Profiles of turbulent stresses in channel flow, as computed by direct numerical simulations (DNS) by \cite{lee2015}.  }
\label{all_stresses}
\end{figure}

By all indications, the streamwise mean velocity ${U}$ in the region $y/\delta \lessapprox 0.15$ is a unique function of $y^+$ that is independent of Reynolds number (see, for example, Zagarola \& Smits 1998, and McKeon {\it et al.\/} 2004). \nocite{Zagarola1998, McKeon:2004}  In contrast, the Reynolds stresses in the near-wall region exhibit a significant dependence on Reynolds number, as illustrated in figure~\ref{all_stresses} for channel flow.  Here, $\overline{u^2}$, $\overline{v^2}$ and $\overline{w^2}$ are in the streamwise, wall-normal and spanwise directions, respectively, and $-\overline{uv}$ is the Reynolds shear stress.
 
The behavior of the streamwise component $\overline{{u^2}^+}$ has been a particular focus of attention, especially its peak value $\overline{{u_p^2}^+}$ located at $y^+ \approx 15$. By experiment, \cite{samie2018fully} showed that in a boundary layer for $6123 \le Re_\tau \le 19680$ $\overline{{u_p^2}^+}$ follows a logarithmic variation given by
\begin{equation}
\overline{{u^2_p}^+} = \beta + \alpha \ln(Re_\tau).
\label{up1}
\end{equation}
with $\alpha=0.646$ and $\beta=3.54$.  \cite{lee2015} found a very similar result from DNS of a channel flow (using only the data for $Re_\tau \ge 1000$) with $\alpha=0.642$ and $\beta=3.66$,
very much in line with the result reported by \cite{Lozano2014}, also obtained by DNS of channel flow, who found $\alpha=0.65$ and $\beta=3.63$.  Finally, \cite{pirozzoli2021reynolds_jfm} found $\alpha=0.612$ and $\beta=3.75$ from DNS for pipe flow at $Re_\tau$ up to 6000.  

We now examine the scaling of all four stress components in the near-wall region using DNS for channel flows \citep{lee2015}, pipe flows \citep{pirozzoli2021reynolds_jfm} and boundary layers \citep{Sillero2011,Sillero2013,WuE5292}.  Couette flows were also considered but it turns out that their behavior is very different from the other canonical flows  \citep{pirozzoli2014, lee2017role, lee_moser_2018}, and therefore they will be considered separately in a future study. Earlier work that was focused on channel flow was presented by 
\cite{smits2021hultmark} and \cite{hultmark2021smits}, as referenced by \cite{monkewitz2021asymptotics}.

\section{Taylor Series expansions}

We begin by writing the Taylor series expansions for $u_i$ in the vicinity of the wall.  Instantaneously \citep{Pope2000, bewley2004skin},
\begin{eqnarray}
\underline u^+ & = & \underline a_1 + \underline b_1 y^+  + \underline c_1{y^+}^2 + \underline d_1{y^+}^3+ O({y^+}^4) \label{Taylor1u} \\
\underline v^+ & = & \underline a_2 + \underline b_2 y^+ + \underline c_2{y^+}^2 + \underline d_2{y^+}^3+ O({y^+}^4) \label{Taylor1v} \\
\underline w^+ & = & \underline a_3 + \underline b_3 y^+ + \underline c_3{y^+}^2 + \underline d_3{y^+}^3+ O({y^+}^4) \label{Taylor1w}
\end{eqnarray}
where $\underline u=U+u$, etc.  The no-slip condition gives $\underline a_1 = \underline a_2 = \underline a_3=0$, and by continuity $\partial \underline v /\partial y|_w = \underline b_2 = 0$.
Also 
\begin{eqnarray}
\underline b_1 & = &  (\partial \underline  u^+/ \partial y^+)_w,  \label{b1}  \\
\underline b_3 & = &  (\partial \underline w^+/ \partial y^+)_w,  \label{b3} \\
\underline c_2 & = & {\textstyle \frac{1}{2}}  (\partial^2 \underline v^+/ \partial {y^+}^2)_w  = -{\textstyle \frac{1}{2}}  (\partial \underline b_1/ \partial x^+ + \partial \underline b_3/ \partial z^+).   \label{c2}
\end{eqnarray}

\noindent
For the corresponding time-averaged quantities
\begin{eqnarray}
  {\overline{{u^2}^+}}/{{y^+}^2} &= F_{u^2} &= f_{u^2} + 2\overline{b_1 c_1} \, y^+            + O({y^+}^2) \label{def_f_uu}\\
  {\overline{{v^2}^+}}/{{y^+}^4} &= F_{v^2} &= f_{v^2} + 2\overline{c_2 d_2} \, y^+            + O({y^+}^2) \label{def_f_vv}\\
  {\overline{{w^2}^+}}/{{y^+}^2} &= F_{w^2} &= f_{w^2} + 2\overline{b_3 c_3} \, y^+            + O({y^+}^2) \label{def_f_ww}\\
  {\overline{uv^{+}}}/{{y^+}^3}  &= F_{uv}  &= f_{uv}  + (\overline{b_1 d_2} + \overline{c_1 c_2}) \, y^+ + O({y^+}^2) \label{def_f_uv}
\end{eqnarray}
where we use the notation $f_{u^2} = \overline{b_1^2}$,  $f_{v^2} = \overline{c_2^2}$,  $f_{w^2} = \overline{b_3^2}$,  and $f_{uv} = \overline{b_1c_2}$.  

\begin{figure}
\centering
\includegraphics[width=\textwidth]{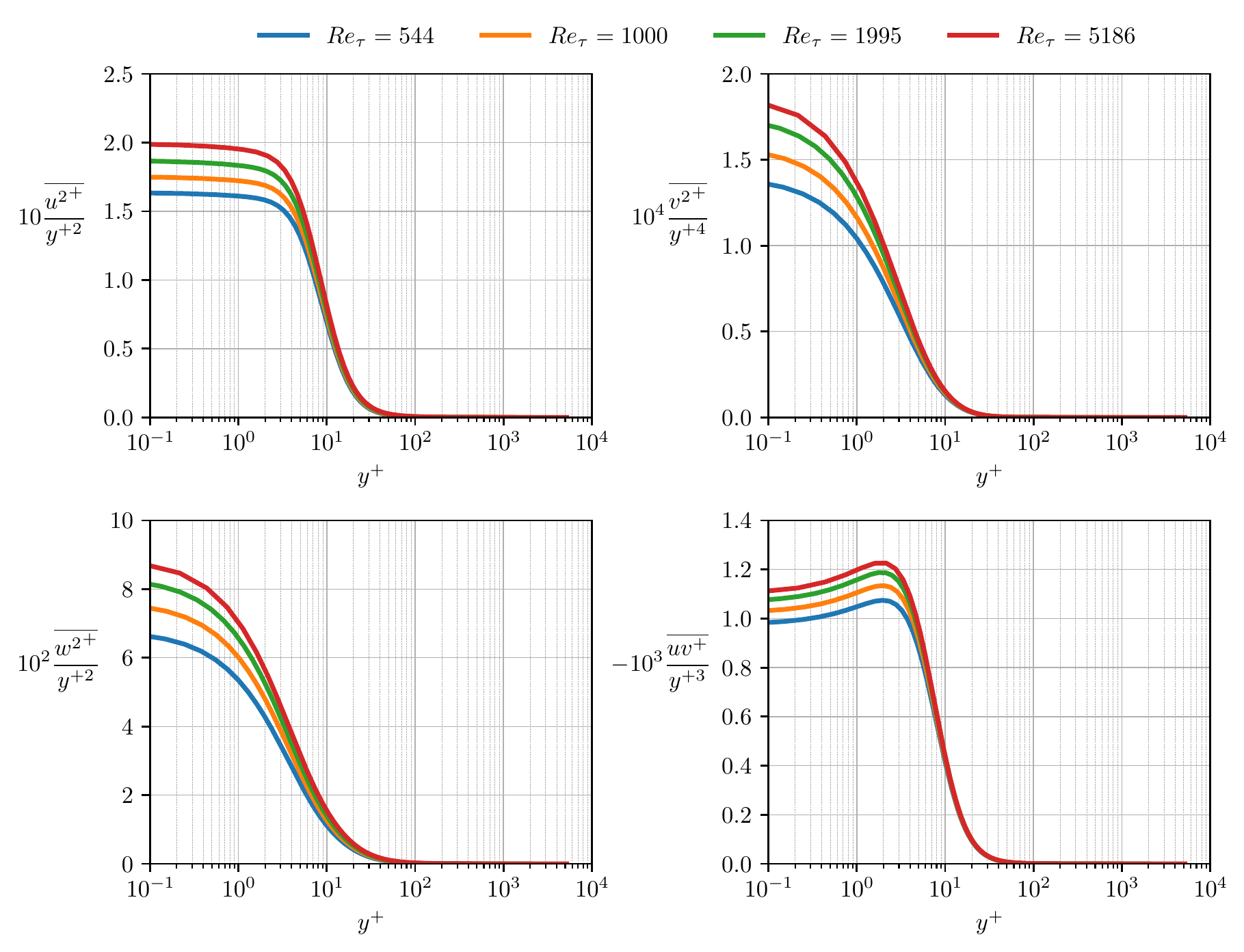}
\caption{ Top left: Profiles of $\overline{{u^{+}}^2}/{{y^+}^2}$.  Top right: $\overline{{v^{+}}^2}/{{y^+}^4}$.  Bottom left: $\overline{{w^{+}}^2}/{{y^+}^2}$.  Bottom right:  $(-\overline{uv^+})/{y^+}^3$. From DNS of channel flow \citep{lee2015}. }
\label{u_U_DNS}
\end{figure}

These functions all become constant as the wall is approached, and the values of $f_{u^2}$, $f_{v^2}$, $f_{w^2}$ and $f_{uv}$ are given by their intercepts at $y^+=0$, as illustrated in figure~\ref{u_U_DNS} for channel flow. Similar results are obtained for the other flows, and they are listed in table~\ref{all_constants}. 

Computing the functions using the definitions in (\ref{b1}--\ref{c2}) requires expensive computations that depend on the dataset. Instead, we use (\ref{def_f_uu}--\ref{def_f_uv}) and employ the Richardson extrapolation method, where the linear approximation of $f_{u^2}$ ($=\widetilde{f}_{u^2}$) is given by
\begin{equation}
f_{u^2} \approx \widetilde{f}_{u^2} = \displaystyle \left.F_{u^2}\right|_{y = y_1^+} - \left( \left.F_{u^2}\right|_{y = y_2^+} - \left.F_{u^2}\right|_{y = y_1^+} \right) \frac{y_1^+}{{y_2^+ - y_1^+}}
\end{equation}
where $y_1^+$ and $y_2^+$ are the distances from the wall of the first two data points. The estimated error is given by the magnitude of the second term on the right.  For the channel flow  the differences between results from the direct computation or the Richardson extrapolation are smaller than the estimated error bounds. 


\begin{table}
\centering
\def~{\hphantom{0}}
\begin{tabular}{lcccccc}
 &  \hspace{2mm} $Re_\tau$ \hspace{2mm} & \hspace{2mm} $10f_{u^2} $ \hspace{2mm} & \hspace{2mm}   $10^4 f_{v^2} $  \hspace{2mm} & \hspace{2mm}    $ 10^2 f_{w^2} $  \hspace{2mm}  &  \hspace{2mm} $-10^3 f^2_{uv} $  \hspace{2mm}   \\[1mm]
Channel                           &~544 &1.64(0.00) &1.40(0.00) &6.79(0.01) &0.95(0.00)\\
\citep{lee2015}                   &1000 &1.75(0.00) &1.58(0.00) &7.64(0.01) &1.02(0.00)\\
                                  &1995 &1.87(0.00) &1.76(0.00) &8.35(0.01) &1.07(0.00)\\
                                  &5186 &1.99(0.01) &1.88(0.05) &8.92(0.15) &1.10(0.01)\\[2mm]
Pipe                              &~495 &1.54(0.01) &1.34(0.12) &6.26(0.25) &1.01(0.00)\\
\citep{pirozzoli2021reynolds_jfm} &1137 &1.74(0.01) &1.69(0.17) &7.63(0.32) &1.13(0.00)\\
                                  &1976 &1.85(0.01) &1.74(0.14) &8.04(0.28) &1.21(0.00)\\
                                  &3028 &1.92(0.01) &1.76(0.12) &8.26(0.27) &1.25(0.01)\\
                                  &6022 &2.02(0.01) &1.86(0.14) &8.70(0.29) &1.28(0.01)\\[2mm]
Boundary layer                    &~500 &1.79(0.01) &1.75(0.30) &7.49(0.55) &    -     \\
\citep{WuE5292}                   &1000 &1.81(0.01) &1.85(0.30) &8.12(0.54) &    -     \\[2mm]
Boundary layer                    &~578 &1.70(0.01) &1.68(0.13) &7.77(0.44) &1.12(0.01)\\
\citep{Sillero2011}               &1307 &1.79(0.01) &1.62(0.12) &8.05(0.42) &1.10(0.02)\\
                                  &1989 &1.87(0.01) &1.68(0.12) &8.36(0.41) &1.12(0.02)\\[2mm]
\end{tabular} 

\caption{Flow and Reynolds number dependence of the functions $f$.  Data only for $Re_\tau > 200$. Estimated uncertainty in parentheses. For the data by \cite{WuE5292}, $f_{uv}$ could not be retrieved due to limited resolution very close to the wall.}
\label{all_constants}
\end{table} 

\begin{figure}
\centering
\includegraphics[width=\textwidth]{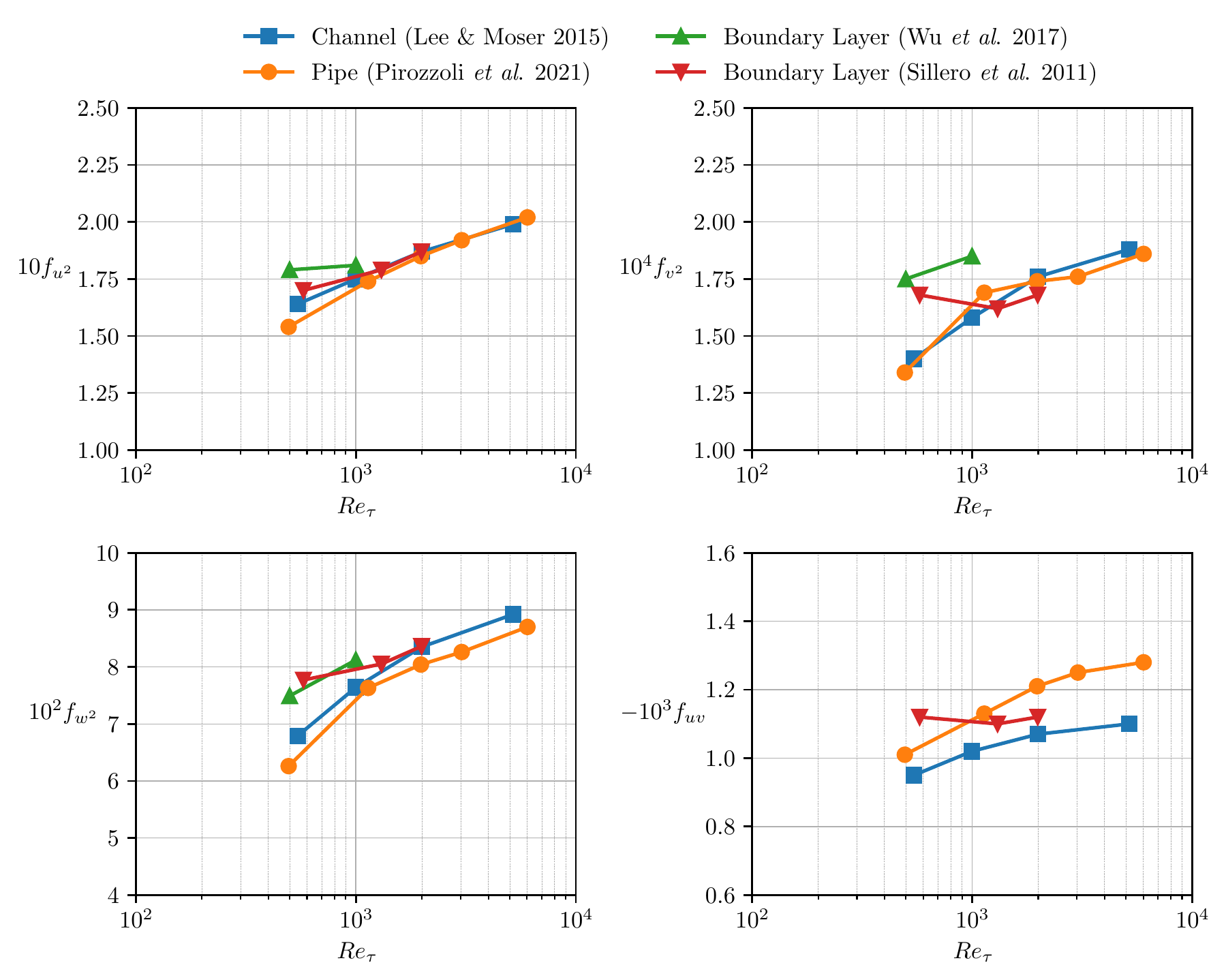}
\caption{Reynolds number variations of functions $f$.  Top left:  $f_{u^2}$.  Top right: $10^3 f_{v^2}$. Bottom left:  $f_{w^2}$.  Bottom right: $10^2 f_{uv}$. $\Box$, channel \citep{lee2015}; {\Large $\circ$}, pipe \citep{pirozzoli2021reynolds_jfm}; $\bigtriangleup$, boundary layer \citep{WuE5292}; $\bigtriangledown$, boundary layer \citep{Sillero2011}. Lines for visual aid only.  The values with uncertainty estimates are listed in table~\ref{all_constants}. }
\label{constants1}
\end{figure}

The Reynolds number dependencies of the functions $f$ are shown in figure~\ref{constants1}.  They all increase with Reynolds number, but at a given Reynolds number there are differences among the values for channel, pipe and boundary layer flows. These trends will be discussed further in \S~4.

\section{Scaling the streamwise stress profiles}

\begin{figure}
\centering
\includegraphics[width=\textwidth]{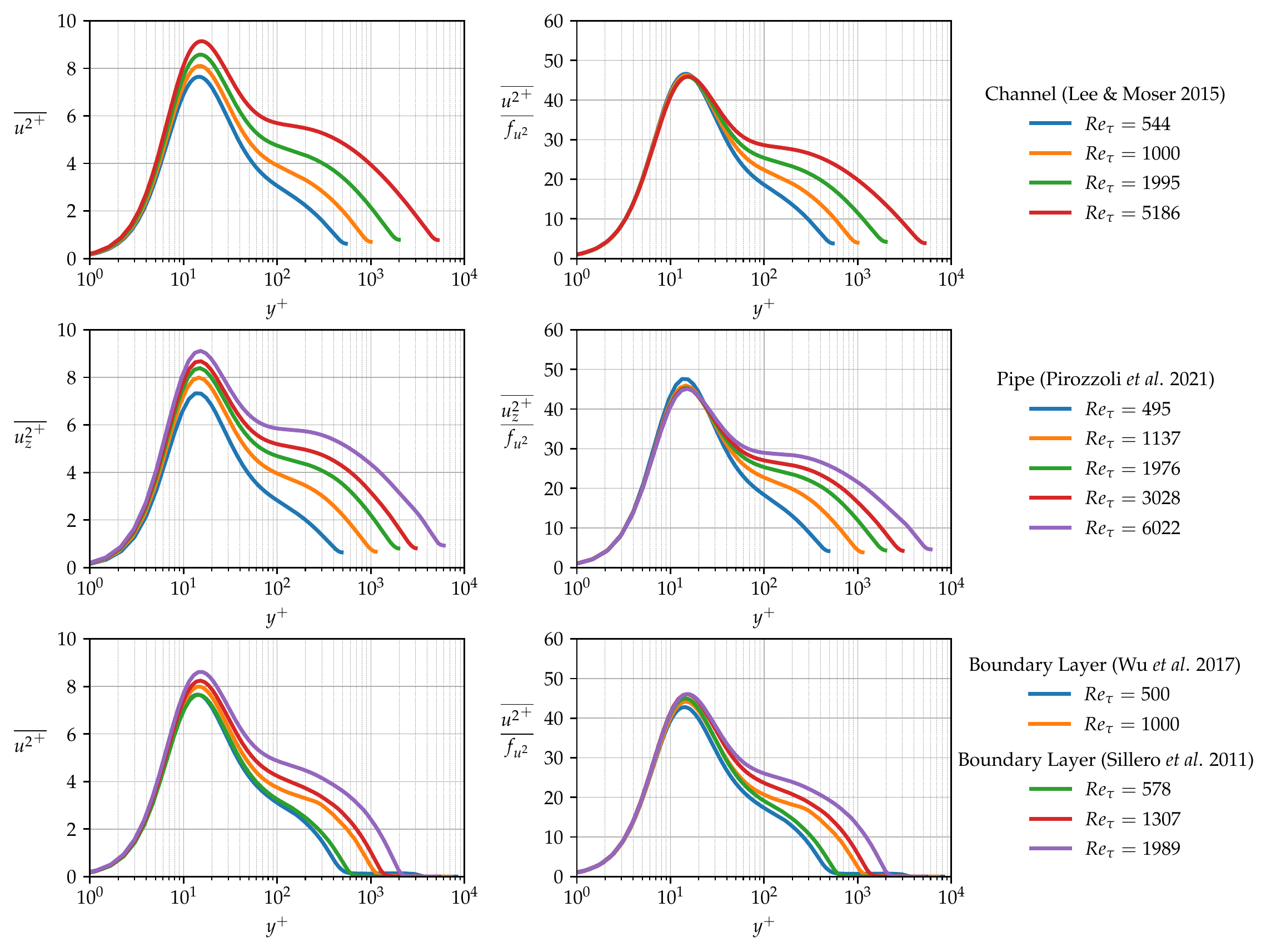}
\caption{ Profiles of streamwise stresses. Left column:  conventional scaling.  Right column:  $f$-scaling. Top row: channel flow \citep{lee2015}. Second row: pipe flow \citep{pirozzoli2021reynolds_jfm}. Third row: boundary layer flow \citep{WuE5292,Sillero2011}.
}
\label{streamwise_stresses}
\end{figure}

Scaling each $\overline{{u^+}^2} $ profile with the value of $f_{u^2}$ at the same Reynolds number yields the results shown in figure~\ref{streamwise_stresses}.  For all three flows, the collapse of the data for $y^+<20$ is impressive, including the almost exact agreement on the scaled inner peak value.  In fact, from table~\ref{inner_peak} it is evident that with increasing Reynolds number $f_{u^2}$ and $\overline{{u_p^+}^2}$ for all three flows approach a constant ratio to each other such that
\begin{equation}
\overline{{u_p^+}^2} \approx 46 f_{u^2}.
\label{u2b12}
\end{equation} 
A similar observation was made previously by \cite{agostini2018impact} and \cite{chen2021reynolds}.  In other words, the magnitude of the peak at $y^+ \approx 15$ tracks almost precisely with $f_{u^2}$, a quantity that is evaluated at $y^+=0$.  

\begin{table}
\centering
\def~{\hphantom{0}}
  \resizebox{\textwidth}{!}{
\begin{tabular}{lcccccccccccc}
&\hspace{0mm} $Re_\tau$ \hspace{0mm} & \hspace{0mm}    $\overline{{u^2_p}^+}$  \hspace{0mm}  & \hspace{0mm} $y^+_{up}$  \hspace{0mm}  &  \hspace{0mm} $\displaystyle \frac{{\overline{{u^2_p}^+}}}{f_{u^2}}$  \hspace{0mm} & \hspace{0mm}  $\displaystyle \frac{\overline{u_p^2}}{u_\tau U_{b}}$  \hspace{0mm}  & \hspace{0mm}  $\displaystyle \frac{\overline{u_p^2}}{u_\tau U_{e}}$  \hspace{0mm}  & \hspace{0mm} $\overline{{w^2_p}^+}$ \hspace{0mm} & \hspace{0mm}  $\displaystyle \frac{\overline{{w^2_p}^+}}{f_{w^2}}$  \hspace{0mm} & \hspace{0mm} $-\overline{uv^+_p}$ \hspace{0mm} & \hspace{0mm}  $(-\overline{uv^+_p})_s$  \hspace{0mm} &  \hspace{0mm} $\overline{{v^2_p}^+}$  \hspace{0mm} &  \hspace{0mm} $(\overline{{v^2_p}^+})_s$  \hspace{0mm}  \\[4mm]
Channel          &~~544 &7.65 &14.8 &46.6 &0.416 &0.364 &1.87 &27.5 &0.862 &1.00 &1.07 &1.24\\
\citep{lee2015}  &~1000 &8.10 &15.2 &46.3 &0.405 &0.359 &2.14 &28.0 &0.903 &1.01 &1.17 &1.30\\
                 &~2003 &8.58 &15.3 &45.9 &0.394 &0.352 &2.42 &29.0 &0.932 &1.00 &1.23 &1.33\\
                 &~5186 &9.14 &15.7 &45.9 &0.379 &0.344 &2.76 &30.9 &0.956 &1.00 &1.28 &1.34\\[2mm]
Pipe                              &~~495 &7.33 &14.3 &47.6 &0.427 &0.336 &1.78 &28.5 &0.852 &1.00 &1.05 &1.23\\
\citep{pirozzoli2021reynolds_jfm} &~1137 &7.99 &14.7 &45.9 &0.413 &0.332 &2.15 &28.2 &0.907 &1.00 &1.18 &1.31\\
                                  &~1976 &8.39 &14.8 &45.4 &0.402 &0.328 &2.37 &29.4 &0.931 &1.00 &1.23 &1.33\\
                                  &~3028 &8.68 &14.9 &45.2 &0.395 &0.328 &2.52 &30.6 &0.944 &1.00 &1.26 &1.34\\
                                  &~6022 &9.11 &15.2 &45.1 &0.385 &0.325 &2.77 &31.8 &0.959 &1.00 &1.29 &1.35\\[2mm]
Boundary layer                    &~~500 &7.66 &14.6 &42.8 &-     &0.349 &1.86 &24.9 &0.917 &1.07 &1.12 &1.31\\
\citep{WuE5292}                   &~1000 &7.99 &15.5 &44.2 &-     &0.322 &2.11 &26.0 &0.934 &1.04 &1.25 &1.39\\[2mm]
Boundary layer                    &~~578 &7.65 &14.4 &45.0 &-     &0.335 &2.06 &26.6 &0.925 &1.07 &1.24 &1.43\\
\citep{Sillero2011}               &~1307 &8.24 &15.2 &46.1 &-     &0.322 &2.38 &29.5 &0.954 &1.05 &1.33 &1.46\\
                                  &~1989 &8.61 &15.9 &46.1 &-     &0.317 &2.56 &30.6 &0.973 &1.05 &1.38 &1.47\\[2mm]
Boundary layer                    &~6123 &9.16 &14.3 &45.5 &-  &0.318 &- &- &-  &-  &- &-  \\
\citep{samie2018fully}            &10100 &9.44 &14.5 &45.3 &-  &0.307 &- &- &-  &-  &- &-   \\
                                  &14680 &9.75 &13.6 &45.7 &- &0.316 &-  &- &-     &- &- &-  \\
                                  &19680 &9.85 &14.8 &45.3 &- &0.305 &-  &- &-     &- &- &-  \\[1mm]
\end{tabular} 
}
\caption{Scaling the inner peak maximum values $\overline{{u^+_p}^2}$, $\overline{{w^+_p}^2}$, $-\overline{uv^+_p}$ and $\overline{{v^+_p}^2}$.  Here, $(-\overline{uv^+_p})_s=-\overline{uv^+_p}/(1-2/\sqrt{\kappa Re_\tau})$, and $(\overline{{v^2}^+_p})_s=\overline{{v^2}^+_p}/(1-2/\sqrt{\kappa Re_\tau})$ with $\kappa= 0.384$. For the boundary layer experiments, $f_{u^2}$ was estimated using equation~\ref{channel_log_f2u}. For the pipe and channels flows, $U_{CL}$ was used instead of $U_e$ in the mixed scaling. }
\label{inner_peak}
\end{table} 

\begin{figure}
\centering
\includegraphics[width=\textwidth]{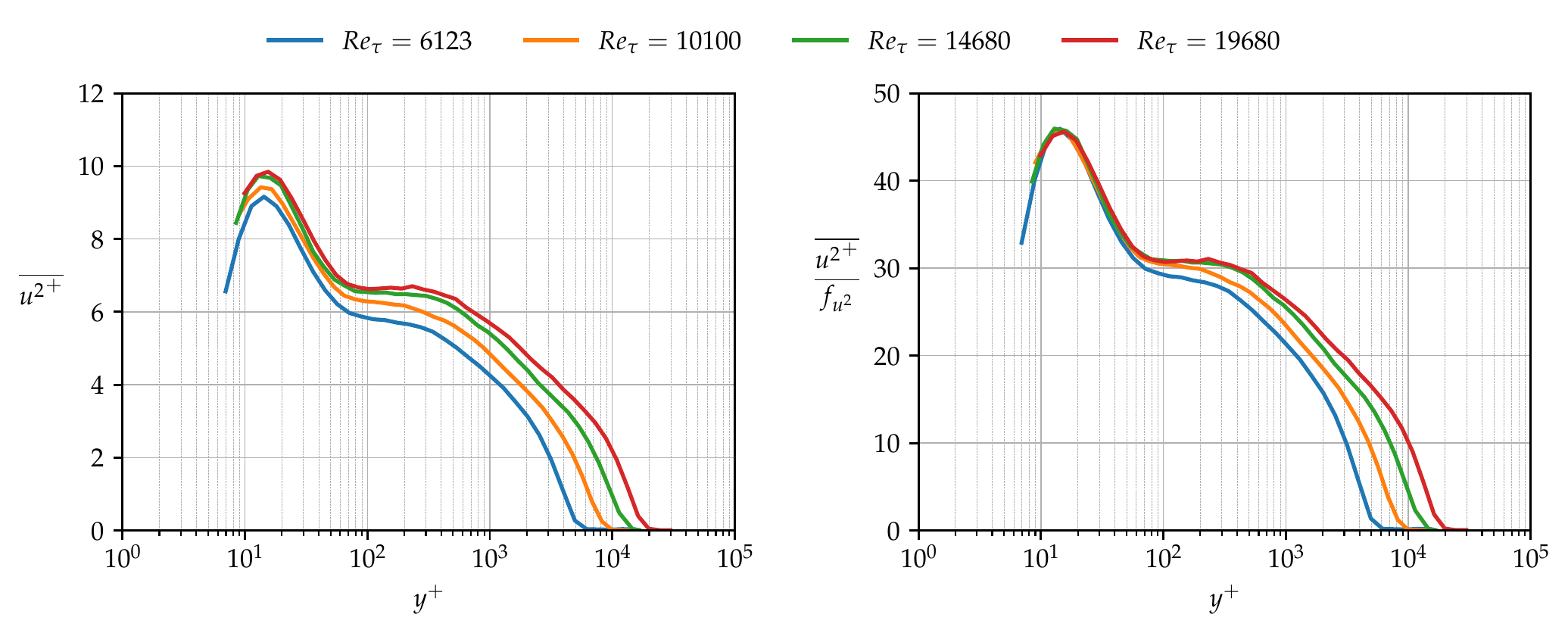} 
\caption{Experimental streamwise stress profiles in boundary layers for $Re_\tau =$ 6,123 to 19,680 \citep{samie2018fully}. Left:  conventional scaling.  Right:  $f$-scaling. }
\label{HiRe_data}
\end{figure}

What about the scaling of $\overline{{u_p^+}^2}$ at higher Reynolds numbers?  Although the collapse of the data shown in figure~\ref{streamwise_stresses} is encouraging, the DNS data only cover a small Reynolds number range.  We can use high Reynolds number experimental data instead, but we need to know what values of $f_{u^2}$ should be used. The highest Reynolds number experiments that are fully resolved are those by \cite{samie2018fully}, but even then data are not available for $y^+<5$, so the values of $f_{u^2}$ cannot be obtained directly from the data.  In this respect, we note that for pipe and channel flow the variation of $f_{u^2}$ for $Re_\tau>1000$ is close to logarithmic, so that 
\begin{equation}
f_{u^2}  =  0.08 + 0.0139 \ln{Re_\tau}.
\label{channel_log_f2u} 
\end{equation}
  For boundary layers a similar relationship appears to fit the data for $Re_\tau>3000$, but the Reynolds number range is too small to make any definite conclusions. If we simply assume that the pipe and channel flow relationship given by equation~\ref{channel_log_f2u} can be used to find the right values of $f_{u^2}$ for high Reynolds number boundary layers, then we obtain the results shown in figure~\ref{HiRe_data}. We see a clear collapse of the data for $y^+<20$, and so it appears that the near-wall profiles of $\overline{{u^2}^+}$ for boundary layers, pipes and channel flows all collapse in this scaling. 

\begin{figure}
\centering
\includegraphics[width=0.55\textwidth]{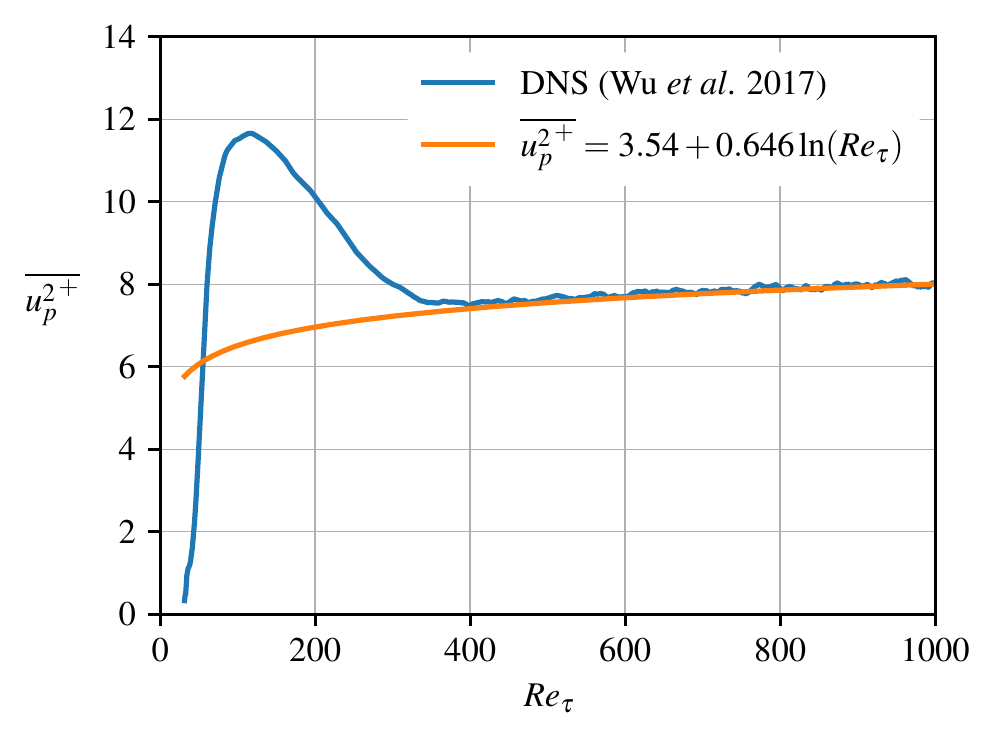}
\caption{Variation of $\overline{u_{p}^{2^{+}}}$ with Reynolds number  in boundary layers. Blue line: DNS \citep{WuE5292}.  Orange line: equation~\ref{up1}  \citep{samie2018fully}.}
\label{DNS_versus_Samie_et_al}
\end{figure}

What about the inverse? Figure~\ref{DNS_versus_Samie_et_al} shows the variation of the peak streamwise turbulence intensity $\overline{u_{p}^{2^{+}}}$ with $Re_{\tau}$ for boundary layers at lower Reynolds numbers. It is seen that the correlation developed by \citep{samie2018fully} based on experimental data from $6000<Re_{\tau}<20000$ agrees very well for the DNS data, at least for $Re_{\tau}>400$. It is thus reasonable to expect the channel flow DNS correlation developed by \cite{lee2015} for $544<Re_{\tau}<5186$ can be used to predict the behavior of $\overline{u_{p}^{2^{+}}}$ in turbulent channel flows at much higher Reynolds numbers.   
 
We can also make some observations on mixed flow scaling.  \cite{DeGraaff2000} proposed that $\overline{u^2}$ in the wall region of boundary layers collapses when scaled with $u_\tau U_e$ versus $y^+$, where $U_e$ is the mean velocity at the edge of the layer (hence the term mixed scaling).  If this is correct, then $\overline{u_p^2}/(u_\tau U_e)$ should be invariant with Reynolds number.  The boundary layer data in table~\ref{inner_peak} supports this proposition at the higher Reynolds numbers.  There is a broad implication here that $f_u^2$ is related in some way to $u_\tau U_e$, but it is not clear what that connection is. 

For pipe and channel flow, to apply mixed scaling we could use either $U_b$ or $U_{CL}$ in the place of $U_e$, where $U_b$ is the bulk velocity and $U_{CL}$ is the centerline velocity. Table~\ref{inner_peak} indicates that the centerline velocity is a better choice, and the high Reynolds number values of $\overline{{u^+_p}^2}/(u_\tau U_{CL})$ for pipe flow are similar to those seen for $\overline{{u^+_p}^2}/(u_\tau U_e)$ in boundary layers.  However, the values of $\overline{{u^+_p}^2}/(u_\tau U_{CL})$ in channel flow are considerably higher than those seen in the other flows.    This behavior is not seen for $\overline{{u^+_p}^2}/f_u^2$ which is more or less constant for all flows, and so the scaling with $f_u^2$ seems to more general than that offered by mixed scaling.  We would not expect universality across different flows for mixed scaling, primarily because $U_{CL}^+$ is flow specific.

Finally, we note that the position of the peak in the streamwise stress, denoted by $y^+_{up}$, is invariant with Reynolds number for all the flows considered here, is shown in table~\ref{inner_peak}.  For all flows, including the high Reynolds number experiments, the value is constant at 15 $\pm 0.8$ (the uncertainty is similar to that set by the resolution of the data in this region).  This value is in accord with most previous estimates.

\section{Scaling the other stress profiles}

\begin{figure}
\centering
\includegraphics[width=\textwidth]{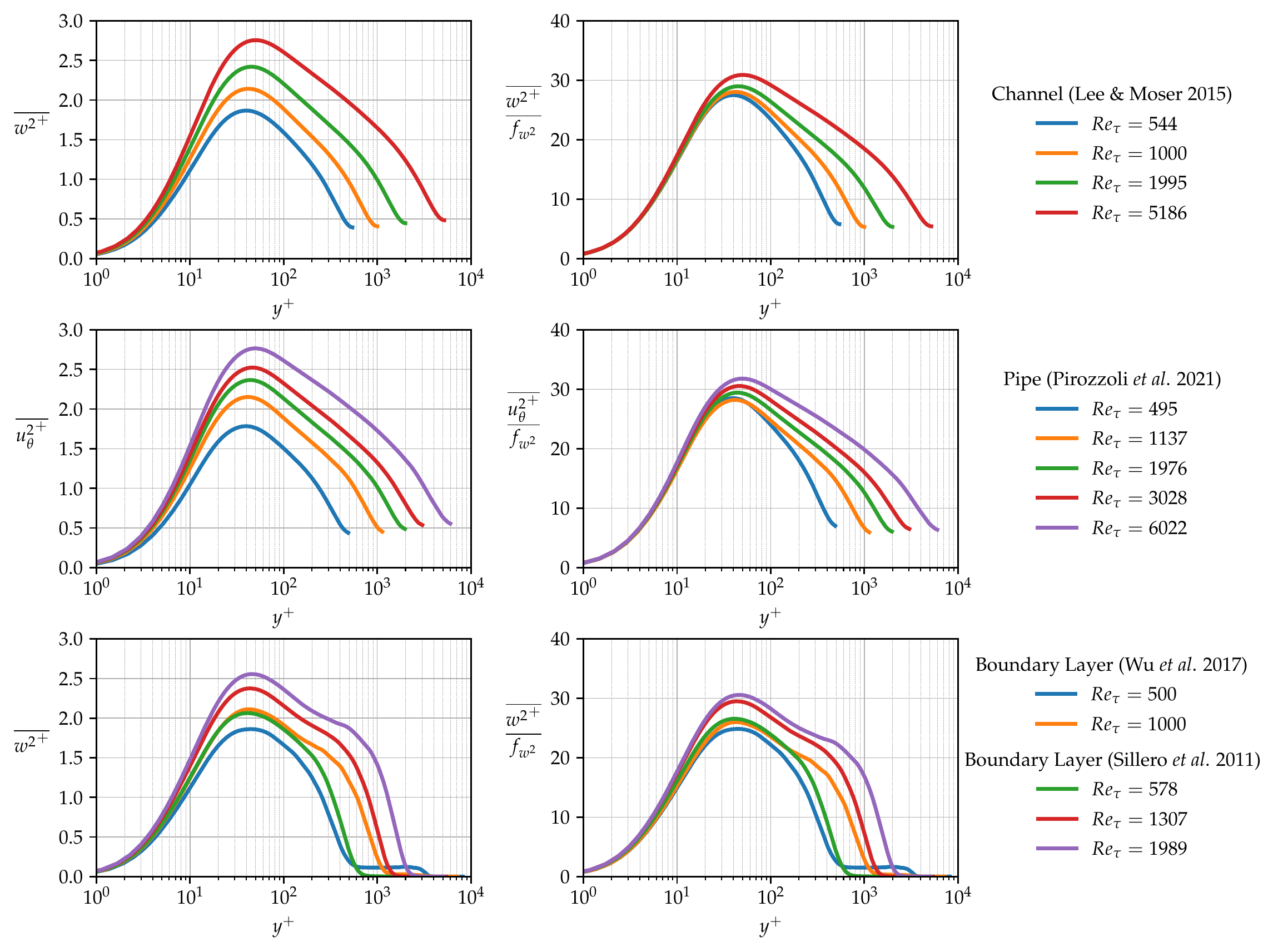} 
\caption{Profiles of spanwise stresses. Left column:  conventional scaling.  Right column:  $f$-scaling.  Top row: channel flow  \citep{lee2015}. Second row: pipe flow  \citep{pirozzoli2021reynolds_jfm}. Third row: boundary layer flow  \citep{WuE5292,Sillero2011}.
}
\label{spanwise_stresses}
\end{figure}

When the $\overline{{w^+}^2} $ profiles are scaled with the value of $f_{w^2}$ at the same Reynolds number, we obtain the results shown in figure~\ref{spanwise_stresses}. The data collapse well for $y^+<20$, but the scaling does not capture the peak value.  From table~\ref{inner_peak} we see that the ratio between $\overline{{w^+_p}^2}$ and $f_{w^2}$ is a slowly increasing function of Reynolds number, and no asymptotic behavior is apparent, at least over this Reynolds number range.  In addition, the location of the peak moves away from the wall with increasing Reynolds number for all three flows.  Thus, for the streamwise and spanwise stresses, the scaling is appropriate only for $y^+<20$ (hence it can capture the peak for $\overline{{u^+}^2} $ but not for $\overline{{w^+}^2}$).


\begin{figure}
\centering
\includegraphics[width=\textwidth]{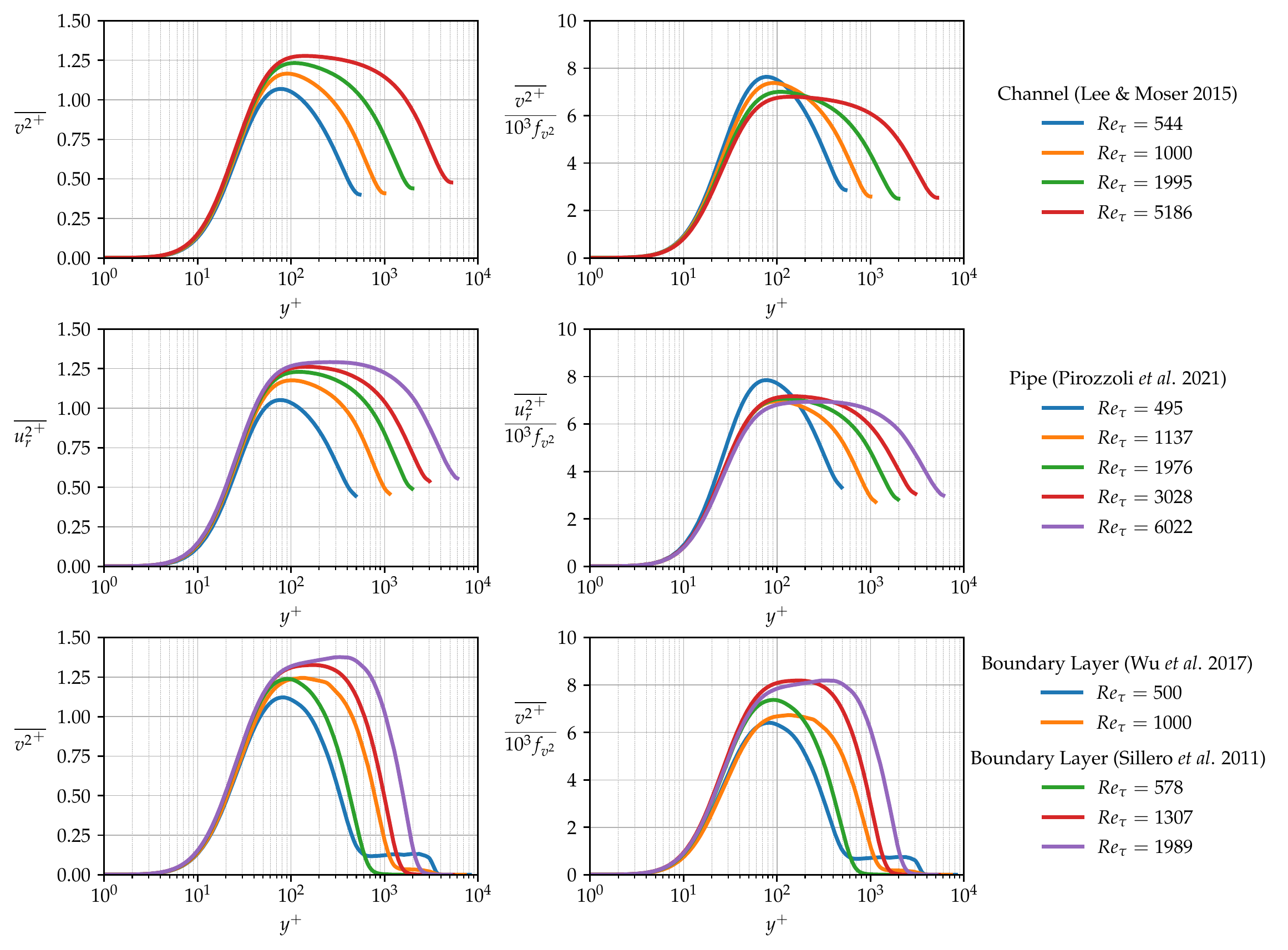}
\caption{ Profiles of wall-normal stresses. Left column:  conventional scaling.  Right column:  $f$-scaling.  Top row: channel flow  \citep{lee2015}. Second row: pipe flow \citep{pirozzoli2021reynolds_jfm}. Third row: boundary layer flow  \citep{WuE5292,Sillero2011}.
}
\label{wallnormal_stresses}
\end{figure}

As for the $\overline{{v^+}^2} $ profiles shown in figure~\ref{wallnormal_stresses}, scaling by $f_{v^2}$ yields modest improvement over the unscaled data, if the lowest Reynolds number cases are disregarded. An explanation for this behavior is advanced in \S~4.   It is clear, however, that the profiles develop a plateau with increasing Reynolds number.  The height of this plateau for the unscaled data may be characterized by the peak value of $\overline{{v^+}^2}$, and these values are given in table~\ref{inner_peak} as $\overline{{v_p^+}^2}$. 

\begin{figure}
\centering
\includegraphics[width=\textwidth]{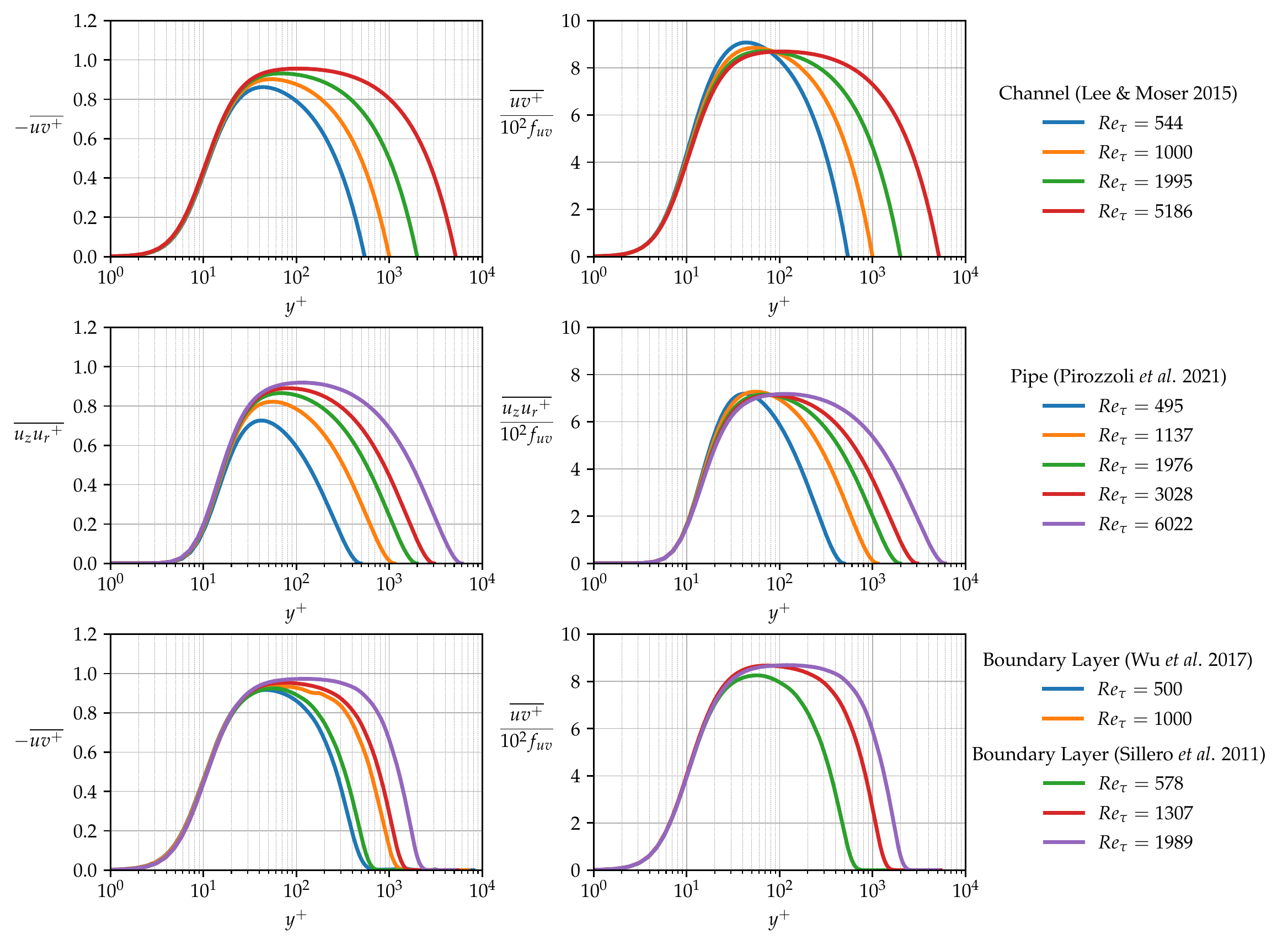}
\caption{ Profiles of shear stresses. Left column:  conventional scaling.  Right column:  $f$-scaling. Top row: channel flow \citep{lee2015}. Second row: pipe flow  \citep{pirozzoli2021reynolds_jfm}.  Third row: boundary layer flow  \citep{WuE5292,Sillero2011}.
}
\label{shear_stresses}
\end{figure}

The shear stress profiles shown in figure~\ref{shear_stresses} display a similar behavior to the normal stress distributions, in that the proposed scaling offers some improvement over the unscaled data, and that a broad plateau appears with increasing Reynolds number.  However, for the pipe and channel flows the extent of the plateau must be bounded at its outer limit by the linear decrease in shear stress dictated by the streamwise pressure gradient.  Its maximum value also cannot exceed one. However, it is rather satisfying to see that the scaling for the peak value proposed for channel flow \citep{lee2015, orlandi2015poiseuille}, that is, $(-\overline{uv^+_p})_s=-\overline{uv^+_p}/(1-2/\sqrt{\kappa Re_\tau})$ with $\kappa= 0.384$, works very well for both channel and pipe flow, as shown in table~\ref{inner_peak}. 

\begin{figure}
\centering
\includegraphics[width=\textwidth]{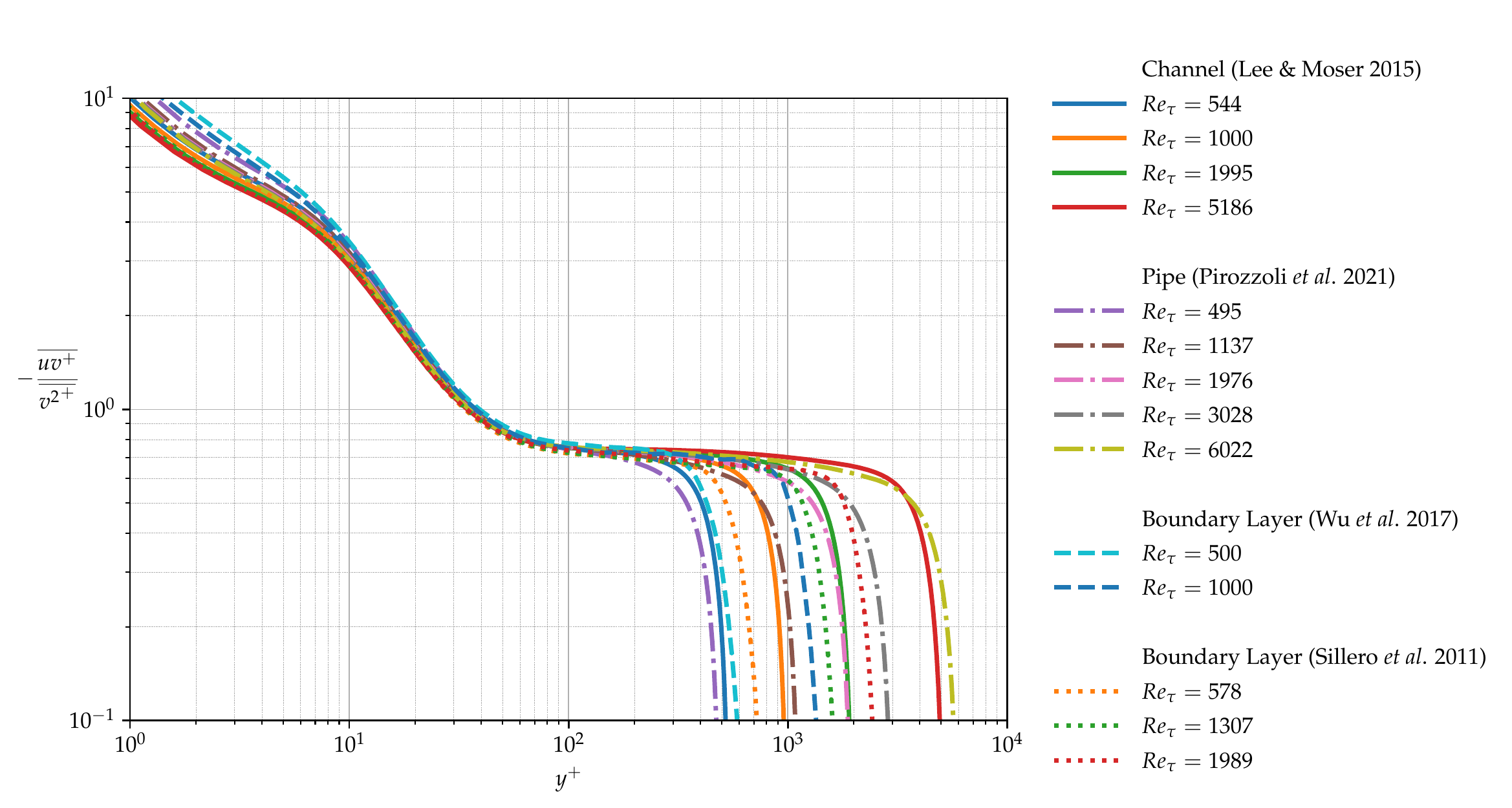} 
\caption{Profiles of $-\overline{uv}/\overline{v^2}$ in channel flow \citep{lee2015}, pipe flow \citep{pirozzoli2021reynolds_jfm} and boundary layer flow \citep{WuE5292,Sillero2011}}
\label{uv_v2}
\end{figure}

We now return to the behavior of the 
peak value of the wall-normal stress, $\overline{{v_p^+}^2}$.  For the channel and pipe flows, the peak level appears to approach a level $>1.3$ with increasing Reynolds number. For $y^+<100$, the ratio $-\overline{uv}/\overline{v^2}$ displays an almost universal behavior, as illustrated for all three flows in figure~\ref{uv_v2}, so that $\overline{{v_p^+}^2}$ should scale like $-\overline{uv_p^+}$.  The values of $(\overline{{v^2}^+_p})_s=\overline{{v^2}^+_p}/(1-2/\sqrt{\kappa Re_\tau})$ are given in table~\ref{inner_peak}, and they indeed tend to approach an asymptotic vale of about 1.38 at high Reynolds number, for both channel and pipe flows.  The asymptotic value for boundary layers appears to be closer to 1.5, although the experiments by \cite{DeGraaff2000} at Reynolds numbers up to 10,070 suggest a level more like 1.4.  This number corresponds to the constant $A_2$ in Townsend's scaling of the wall-normal fluctuations in the logarithmic region, as derived from the attached eddy hypothesis \citep{Townsend1976}.

\section{What does it all mean?}
\label{What}

The Taylor series expansion revealed that
\begin{eqnarray}
f_{u^2} = \overline{b_1^2} & = & \overline{\left( \frac{\partial u^+}{\partial y^+} \right)^2_w} = \frac{\overline{{\tau_{wx}}^2}}{\underline{\tau}_w^2} \\
f_{w^2} = \overline{b_3^2} & = & \overline{\left( \frac{\partial w^+}{\partial y^+} \right)^2_w} = \frac{\overline{{\tau_{wz}}^2}}{\underline{\tau}_w^2}. 
\label{f2u_f2w_def}
\end{eqnarray}
For wall-bounded flows, therefore, the controlling parameter in the near-wall scaling for $u$ is the mean square of the fluctuating wall stress in the $x$-direction $\tau_{wx}$, and for $w$ it is the mean square of the fluctuating wall stress in the $z$-direction $\tau_{wz}$.  
For the other two functions, the Taylor series expansion gives
\begin{eqnarray}
f_{v^2} & = & \overline{c_2^2} = {\frac{1}{4}}  \overline{ \left( \frac{\partial^2 \underline v^+}{ \partial {y^+}^2} \right)^2_w } = -{ \frac{1}{2}} \overline{ \left( \frac{\partial b_1 } {\partial x^+} + \frac{ \partial b_3} {\partial z^+} \right)^2 } \\[1mm] 
f_{uv} & = & \overline {b_1 c_2}
\label{f2v_fuv_def}
\end{eqnarray}
where $b_1$, $b_3$ and $c_2$ are the fluctuating parts of $\underline b_1$, $\underline b_3$ and $\underline c_2$, as given in equations~\ref{b1}-\ref{c2}.  The functions $f_{v^2}$ and $f_{uv}$ therefore  express correlations between spatial gradients of the instantaneous wall stress fluctuations, as well as the fluctuating wall stress itself, and so it is more difficult to give a precise meaning to $f_{v^2}$ and $f_{uv}$, although they are clearly more connected to the small-scale motions than either $f_{u^2}$ or $f_{w^2}$.  Also, by continuity, we see that it is essentially the gradients of the fluctuating shear stress ($b_1$ and $b_2$) that give rise to the wall-normal motion through continuity. 

This connection with the fluctuating wall stress helps to explain the Reynolds number dependence of the functions $f$ shown in figure~\ref{constants1}. With increasing Reynolds number the large-scale (outer layer) motions contribute more and more to the fluctuating wall stress by modulation and superimposition of the near-wall motions \citep{Marusic2010a, orlu2011, mathis2013estimating, agostini2016validity, yang2017, agostini2018impact, lee_jfm2019} (see also \S~6).  Of course, it is the turbulence that controls the wall stress, and not vice versa, but the main point is that the whole of the region $y^+ < 20 $ (including the peak in $\overline{{u^2}^+}$) scales with the velocity scale $u_s=u_\tau \sqrt{f_{u^2}}$, which can be determined by measuring the fluctuating wall stress, a clear indication of the increasingly important contribution of the large-scale motions on the near wall behavior as the Reynolds number increases.  Because $f_{v^2}$ and $f_{uv}$ are more connected to the small-scale motions than either $f_{u^2}$ and $f^2_{w}$, it might be expected that they feel the effects of modulation more than the effects of superimposition by the large-scale motions \citep{Marusic2010a}.

\begin{figure}
\centering
\includegraphics[width=\textwidth]{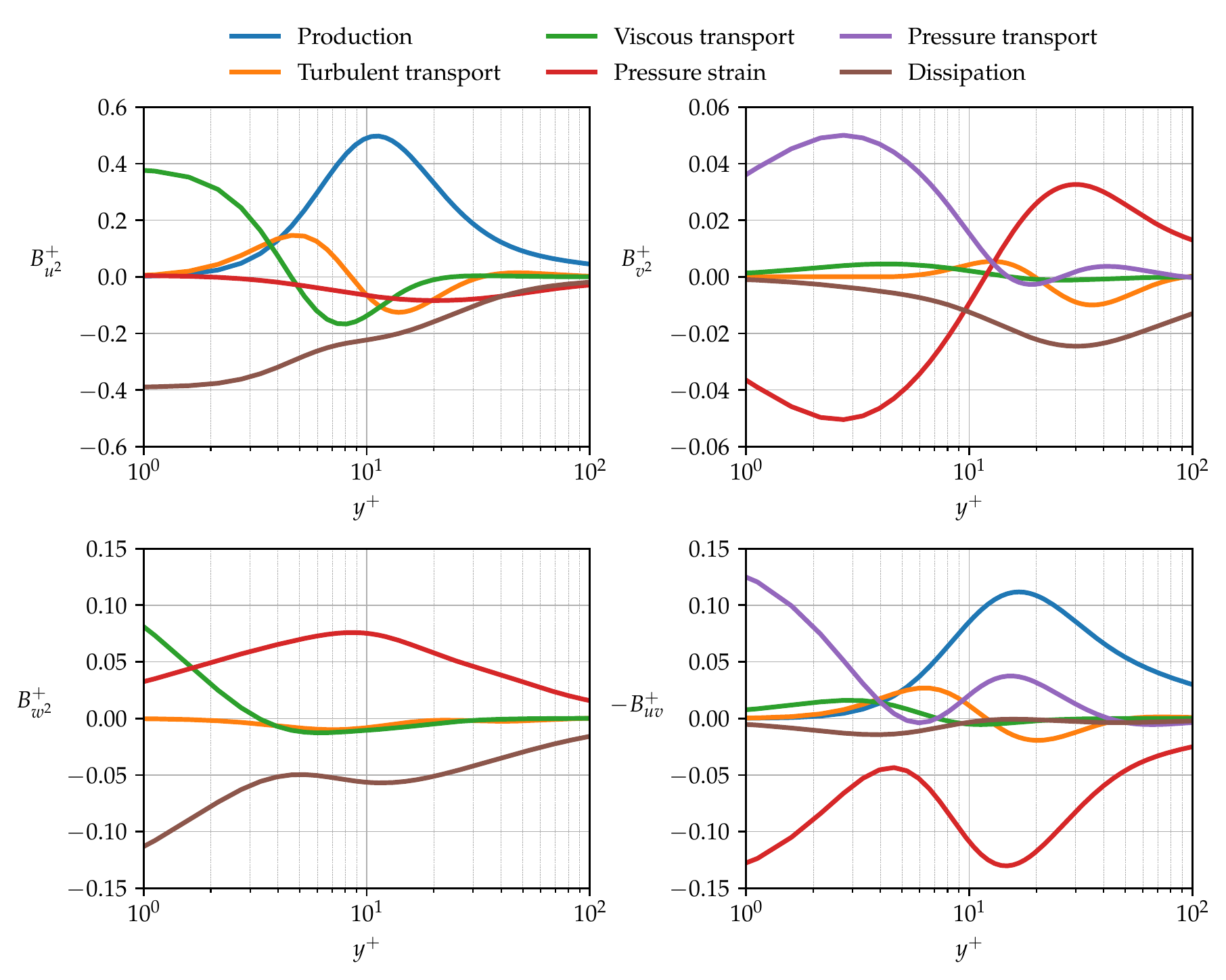} 
\caption{ Energy budgets for turbulent stresses in channel flow   at $Re_\tau=5186$ \citep{lee2015}.  Top left: $\overline{{u^2}^+}$.  Top right: $\overline{{v^2}^+}$.  Bottom left: $\overline{{w^2}^+}$.  Bottom right: $-\overline{{uv}^+}$. }
\label{all_budgets}
\end{figure}

An interpretation based on dissipation scaling rather than wall stress scaling was offered by \cite{chen2021reynolds}.  By using the energy budget for $\overline{{u^2}^+}$, they noted that $\sqrt{f_u^2}$ is equal to the dissipation rate for $\overline{{u^2}^+}$ at the wall, ${\large \mbox{$\epsilon_{u_w}^+$}}$. What's more, close to the wall, the dissipation is balanced by viscous transport, and all other terms are small, as illustrated for channel flow in figure~\ref{all_budgets}.  According to \cite{chen2021reynolds}, this leads to two conclusions.  The first is that the order of the peak value of $\overline{{u^2}^+}$ can be estimated as ${\large \mbox{$\epsilon_{u_w}^+$}} y^+_p$, which yields the same result as that given by equation~\ref{u2b12}.  The second is that, because the dissipation is proposed to be bounded at infinite Reynolds number, the logarithmic increase in ${f_u^2}$ (equation~\ref{channel_log_f2u}), and by extension the logarithmic increase in $\overline{{u_p^2}^+}$ (equation~\ref{up1}), need to be reconsidered. They then suggest an alternative formulation for the peak magnitude that approaches a finite limit at infinite Reynolds number, based on the proposed Reynolds number dependence of the dissipation. 

However, figure~\ref{all_budgets} shows that the energy balance for $\overline{{u^2}^+}$ changes rapidly with distance from the wall, so that by $y^+=10$ the production dominates and the viscous transport has actually changed sign.  It is not clear, therefore, why a dissipation scaling should persist all the way to $y^+=20$. One of the assumptions made by \cite{chen2021reynolds} is that the production balances the dissipation at the location where the production is maximum, which is close to the point where $\overline{u^2}$ has its peak value.  However, \cite{lee2015} showed the largest imbalance of the production and dissipation actually occurs at that particular location.  Our interpretation, based on the wall stress signature, has the benefit of reflecting more directly the influence of the large-scale motions in the near-wall region, and so may offer a more robust explanation for the near-wall scaling.  

The energy budget for $\overline{{w^2}^+}$ also indicates that very close to the wall the viscous transport is balanced by dissipation ${\large \mbox{$\epsilon_{w_w}^+$}}$.  However, by about $y^+ \approx 3$, the balance changes so that now pressure strain balances the dissipation and all other terms are small.  Again. it seems more natural therefore to build a scaling argument on the behavior of the wall stress fluctuations rather than the dissipation. 

Finally, the energy budgets for $\overline{{v^2}^+}$ and $-\overline{{uv}^+}$  very close to the wall indicate that for both stresses  pressure strain is balanced by pressure transport, which must both go to zero at the wall (figure~\ref{all_budgets}). For $\overline{{v^2}^+}$, the balance changes rapidly with distance from the wall so that by $y^+>10$ the pressure strain/dissipation balance dominates. Even more interestingly, for $-\overline{{uv}^+}$ the dissipation is not important anywhere in the near-wall region and for $y^+>10$ the pressure strain/production balance dominates.


\section{Two-dimensional spectral density of $f_{u^2}$, $f_{v^2}$, $f_{w^2}$, $f_{uv}$}

To help understand the influence of the large scale motions on the fluctuations in the wall stress, we now examine the spectral structure of the proposed scaling parameters. The spectral densities of the functions $f$'s are given by 
\begin{eqnarray}
  E_{f_{u^2}}(k_x, k_z) &= 2\mathrm{Re} \{ \overline{ \widehat{b_1} \widehat{b_1^*}} \} \\
  E_{f_{v^2}}(k_x, k_z) &= 2\mathrm{Re} \{ \overline{ \widehat{c_2} \widehat{c_2^*}} \} \\
  E_{f_{w^2}}(k_x, k_z) &= 2\mathrm{Re} \{ \overline{ \widehat{b_3} \widehat{b_3^*}} \} \\
  E_{f_{uv}}(k_x, k_z)  &= 2\mathrm{Re} \{ \overline{ \widehat{b_1} \widehat{c_2^*}} \}
  \label{eq:def_2d_spectra}
\end{eqnarray}
where ${\hat{\cdot}}$ denotes Fourier transformation in the $x$ and $z$ directions, and ${\hat{\cdot}^*}$ denotes the complex conjugate of ${\hat{\cdot}}$. Also, $k_x$ and $k_z$ are the wavenumbers in the $x$ and $z$ directions, respectively. We will use the polar-log coordinate system introduced by \citet{lee_jfm2019} to investigate the spectral structure in terms of length scales and anisotropy. In this approach, the two-dimensional spectral densities in Cartesian coordinates $(k_x, k_z)$ are mapped to the polar-log coordinates $(k_x^\sharp, k_z^\sharp)$ with corresponding Jacobians. For example,
\begin{equation}
  f_{u^2} = \iint E_{f_{u^2}} \mathrm{d} k_x \mathrm{d} k_z = \iint \frac{|k|^2}{\xi} E_{f_{u^2}} \mathrm{d}k_x^\# \mathrm{d}k_z^\#,
\end{equation}
and
\begin{equation}
  k_x^\# = \frac{k_x}{|k|} \xi , \quad k_z^\# = \frac{k_z}{|k|}\xi,
\end{equation}
where
\begin{equation}
|k| = \sqrt{k_z^2 + k_z^2}  \quad\mathrm{and}\quad  \xi = \log_{10} \frac{|k|}{k_\mathrm{ref}},
\end{equation}
where we will choose the reference wavenumber $k_\mathrm{ref} = Re_\tau / 50000$. In this form, the anisotropy of the spectral density and the contributions of $k_x = 0$ and $k_z = 0$ are clearly represented. The more commonly used premultiplied two-dimensional spectral density, $k_x k_z E(\log k_x, \log k_z)$ suppresses the contribution to the spectral density when either $k_x$ or $k_z$ is small, thereby masking the influence of the large scale motions.  More details on the representation of two-dimensional spectral densities in the polar-log coordinate system are given by \cite{lee_jfm2019}. 

\begin{figure}
\centering
\includegraphics[width=0.9\textwidth]{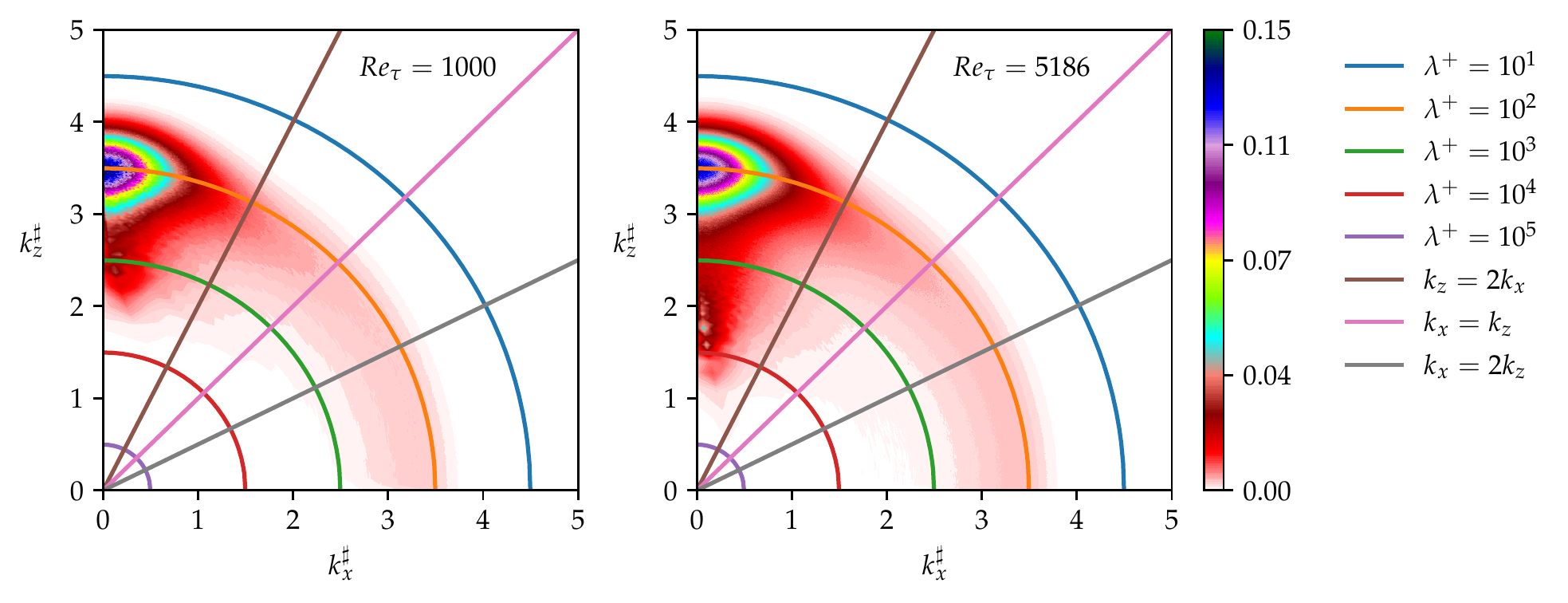}
\caption{Two-dimensional spectral density of $f_{u^2}$ in channel flow, using the DNS by \cite{lee2015}. Left: $Re_\tau = 1000$. Right: $Re_\tau = 5186$}
\label{fig:2d_uu}
\end{figure}

The spectral densities of $f_{u^2}$, shown in figure~\ref{fig:2d_uu}, indicate that streamwise elongated motions ($2k_x < k_z$)  dominate the energy content. Also, motions with $\lambda^+ \ \approx 100$ make the largest contributions, which is consistent with the spectral density of $u'^2$ which has a peak at $\lambda_z^+ = 100$ in near-wall flows, corresponding to the spacing of the near-wall streaks. Furthermore, with increasing Reynolds number the contributions by the large-scale motions increase; compare, for example, the contributions by motions with $\lambda^+ > 1000$, a trend that is consistent with previous work \citep{orlu2011,cimarelli2015sources,lee_jfm2019}.  

\begin{figure}
\centering
\includegraphics[width=0.55\textwidth]{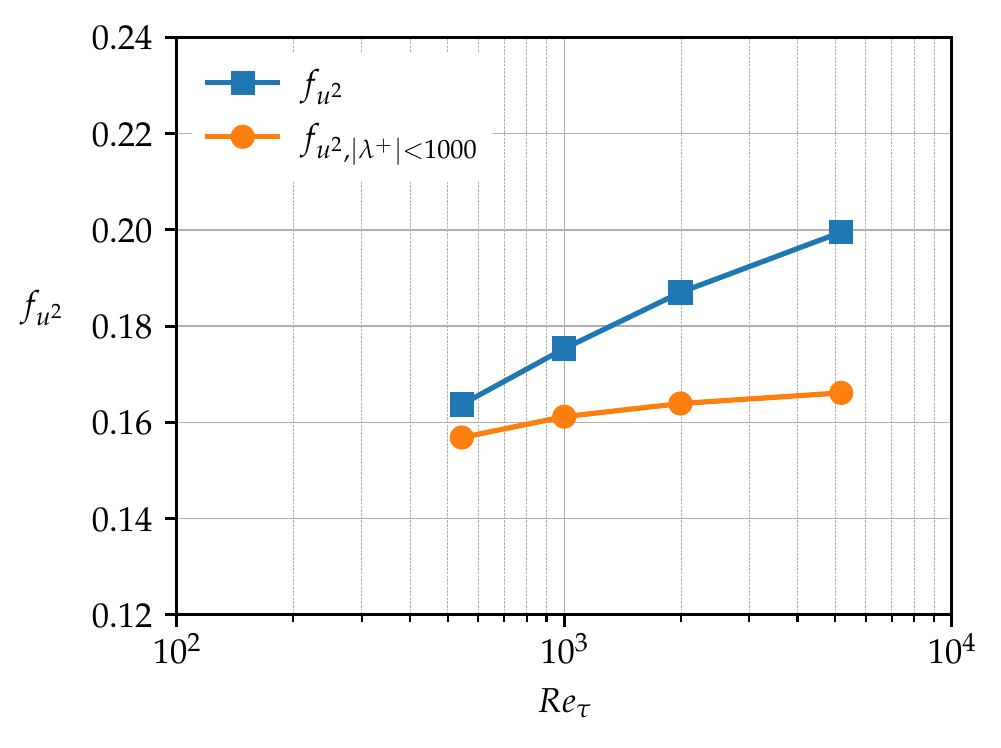}
\caption{Contribution of small-scale motions to $f^2_{u}$ in channel flow, using the DNS by \cite{lee2015}.}
\label{fig:f_uu_filter}
\end{figure}

To quantify the contributions of large-scale motions and small-scale motions to $E_{f_{u^2}}$, we use a high-pass filter according to
\begin{equation}
  f^2_{u,SS} = \int_{|\mathbf{k}|> k_\mathrm{c}} E_{f_{u^2}} \mathrm{d} \mathbf{k},
\end{equation}
where $k_\mathrm{c}$ is the cut-off frequency. For the high Reynolds number case shown in figure~\ref{fig:2d_uu}b, a convenient demarcation between small-scale and large-scale motions occurs at $\lambda^+ \approx 1000$, and so we choose  $k_\mathrm{c}^+ = 2\pi/1000$. The results are given in figure~\ref{fig:f_uu_filter}.
Whereas $f_{u^2}$ increases with Reynolds number, as shown earlier in figure~\ref{constants1}, the small-scale contribution $f^2_{u,SS}$ is almost invariant.  Because the small-scale motions ($\lambda^+ < 1000$) are universal in the near-wall region \citep{lee_jfm2019}, we conclude that $f_{u^2}$ is the correct scaling parameter for near-wall flows at high Reynolds number since it properly measures the contributions by large-scale motions.

\begin{figure}
\centering
\includegraphics[width=0.9\textwidth]{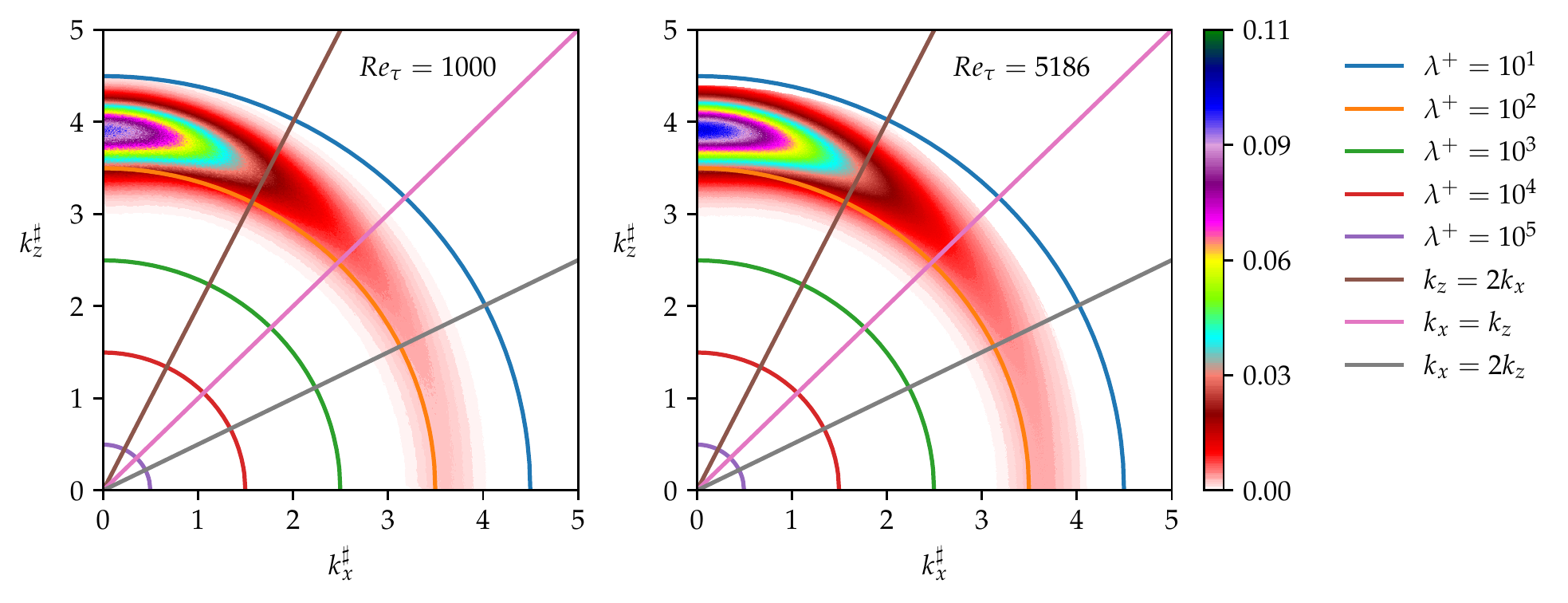}
\includegraphics[width=0.9\textwidth]{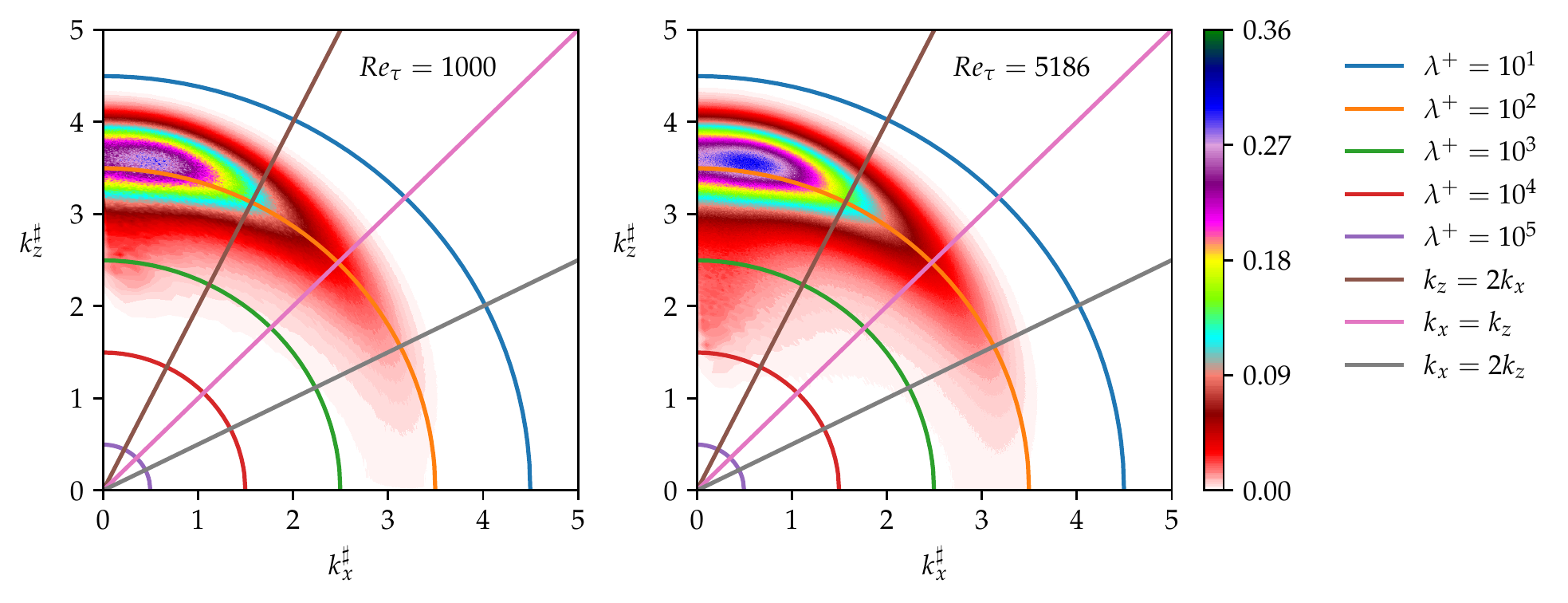}
\includegraphics[width=0.9\textwidth]{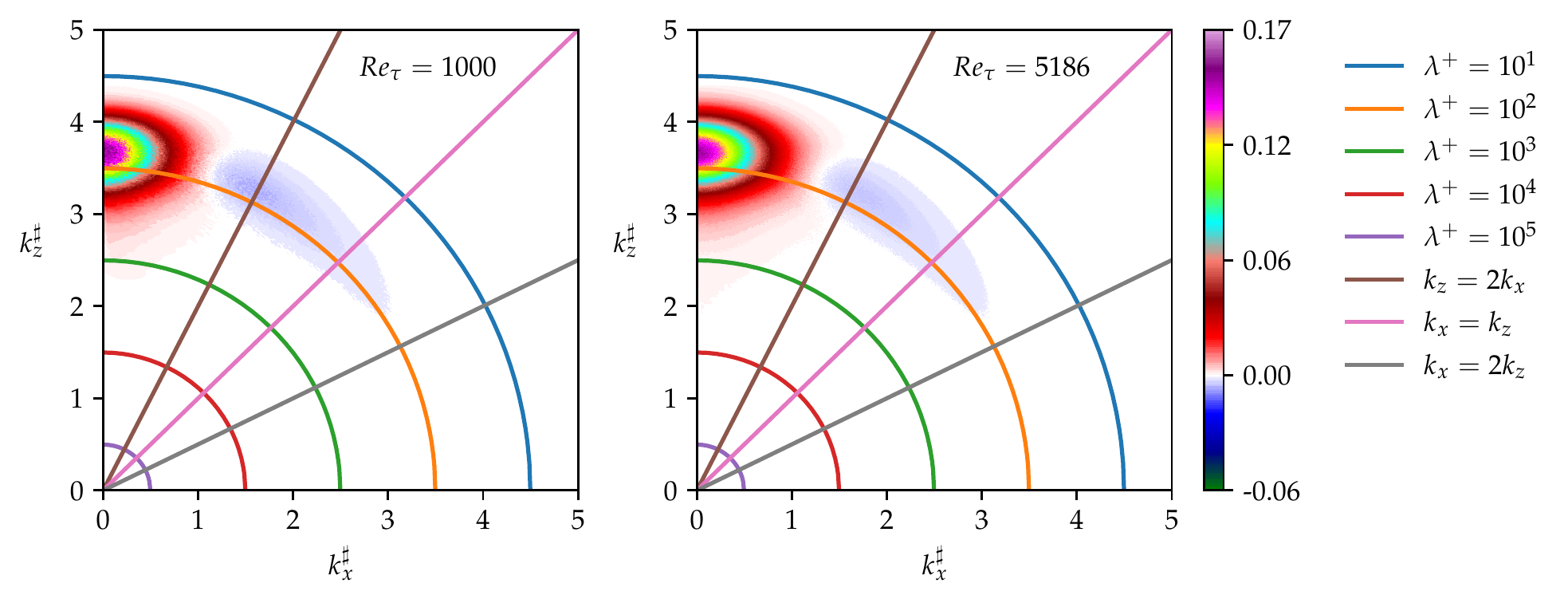}
\caption{2D spectral density of $f$s in channel flow  \citep{lee2015}. Left: $Re_\tau = 1000$. Right: $Re_\tau = 5186$. Top row: $10^3f_{v^2}$.  Middle row: $10f_{w^2}$.  Bottom row: $-10^2f_{uv}$.}
\label{fig:2d_others}
\end{figure}

The spectral densities of $f_{v^2}$, $f_{w^2}$ and $f_{uv}$ are shown in figure~\ref{fig:2d_others}. Only $f_{w^2}$ shows an increasing contribution of large-scale motions as the Reynolds number increases but it is weak relative to what was seen for $f_{u^2}$. The peak values of both $f_{v^2}$ and $f_{w^2}$ increase Reynolds number at fixed length scales, but the peak in $E_{f_{w^2}}$ occurs at a smaller length scale than the peak in $E_{f_{u^2}}$ and the peak of $E_{f_{v^2}}$ is at an even smaller length scale, independent of Reynolds number. Also, $E_{f_{u^2}}$ and $E_{f_{w^2}}$ show a streamwise-elongated structure, whereas $E_{f_{vv}}$ has a more isotropic structure.  Unique among these functions, $f_{uv}$ has a negative contribution where $2 k_x > k_z$. Similar features are seen in the  spectral densities of $-u'v'$ \citep{lee_jfm2019}, but the underlying mechanism of this negative contribution is not clear. 

\section{Conclusions}

By expanding the velocity in a Taylor series with distance from the wall, the Reynolds number dependence of the near-wall distributions of the Reynolds stresses was traced to the magnitude of the fluctuating wall shear stress and its spatial gradients, which are increasingly affected by the superimposition and modulation of the near-wall motions due to large-scale, outer-layer motions as the Reynolds number increases.  

The Taylor series expansion also suggests a separate scaling for each component of the Reynolds stress.  For the streamwise and spanwise components, the scaling collapses the data for $y^+ < 20$, a region that includes the near-wall peak in $\overline{{u^2}^+}$ but not the one in $\overline{{w^2}^+}$. For the wall-normal component and the Reynolds shear stress, the proposed scaling offers a modest improvement in the collapse of the data over traditional scaling, one that appears to get better at higher Reynolds numbers. 

Revisiting the dimensional analysis given in equation~\ref{dim_analysis}, we can now be more precise and write for the Reynolds stresses in a two-dimensional wall-bounded flow, for the region $y^+<20$,
\begin{equation}
\overline{(u_i u_j)^+} = f(Re_\tau) g(y^+).
\label{dim_analysis2}    
\end{equation}
That is, it is possible to separate the dependence on Reynolds number from the dependence on wall distance.  

It may also be remarked that because the scaling is different for each component of the stress, any isotropic definition of eddy viscosity will obviously fail in the near-wall region.  In this respect, \cite{Hultmark2013a} noted that in the overlap region, where both the mean velocity $U$ and the streamwise stress $\overline{{u^2}^+}$ follow a logarithmic distribution, $\overline{{u^2}^+}$ depends on $U$ rather than its gradient.

More generally, there are many similarities among the three flows considered here.  In particular, within the uncertainty limits, the values for $f_{u^2}$, $f_{v^2}$, $f_{w^2}$ are similar for all three flows, at least for $Re_\tau > 1000$, at which point they show almost identical growth with further increases in Reynolds number. As to $f_{uv}$, the differences between pipe and channel flows grow with Reynolds number. A contributing factor may be that the large-scale motions are subject to different geometric constraints. In terms of the energy budget, the key factor appears to be the pressure, in that pressure strain and transport are the dominant terms in the budget of $\overline{uv}$.  

The authors perceive no conflict of interest in submitting this manuscript.

\subsection*{Acknowledgments}
The authors would like to thank Matt Fu and Liuyang Ding for comments on an earlier draft.  This work was supported by ONR under Grant N00014-17-1-2309 (Program Manager Peter Chang). SP acknowledges that the pipe flow DNS results reported in this paper have been achieved using the PRACE Research Infrastructure resource MARCONI based at CINECA, Casalecchio di Reno, Italy, under project PRACE n. 2019204979.  An award of computer time was provided by the Innovative and Novel Computational Impact on Theory and Experiment (INCITE) program. This research used resources of the Argonne Leadership Computing Facility, which is a DOE Office of Science User Facility supported under Contract DE-AC02-06CH11357.

Sandia National Laboratories is a multimission laboratory managed and operated by National Technology and Engineering Solutions of Sandia, LLC., a wholly owned subsidiary of Honeywell International, Inc., for the U.S. Department of Energy's National Nuclear Security Administration under contract DE-NA0003525. This paper describes objective technical results and analysis. Any subjective views or opinions that might be expressed in the paper do not necessarily represent the views of the U.S. Department of Energy or the United States Government.

\bibliographystyle{jfm}
\bibliography{BigBib_master_3_19_21}

\end{document}